\documentclass[12pt]{article}
\usepackage{a4,amsmath,epsfig,amssymb}
\usepackage[latin1]{inputenc}
\usepackage{graphicx}
\usepackage{caption}
\usepackage{subcaption}
\usepackage{feynmp}
\begin{document}
\begin{titlepage}
\renewcommand{\thefootnote}{\fnsymbol{footnote}}
\vspace{1.0em}

\begin{center}
{\bf \Large{Phenomenological Analysis 
\\
of the Decay $B^{\pm}\to K^{\pm} p {\bar p}$}}
\end{center}
\vspace{1.0em}
\begin{center}
\begin{large}

E.~Di~Salvo$^{1,2,3}$\footnote{Elvio.Disalvo@ge.infn.it} and
F.~Fontanelli$^{1,2}$\footnote{Flavio.Fontanelli@ge.infn.it}

\end{large}
\bigskip

$^{1}$ Dipartimento di Fisica \\
Via Dodecaneso 33 \\
16146 Genova - ITALY 

\noindent

$^{2}$ INFN - Sez. di Genova. ITALY

\noindent

$^{3}$ Laboratoire de Physique Corpusculaire de Clermont-Ferrand \\
IN2P3/CNRS Universit\'e Blaise Pascal \\
F-63177 Aubi\`ere Cedex. FRANCE \\ 
\end{center}
\vspace{5.0em}
\begin{abstract}
\vspace{1.0em}
\noindent

We propose a parametrization for interpreting some of the presently
 available data of the $B^{\pm} \to K^{\pm} p {\bar p}$ decay, 
 in particular those by LHCb and Belle collaborations. 
 The model is inspired by the well-known current and transition 
 contributions, usually assumed in this kind of decay. However, in 
 the light of considerations about the dominant diagrams and about final   state interactions, we modify some parameters of the model, determining them  by means of a best fit to data. We show the results, which we discuss in some detail. Moreover we give some predictions on other observables relative to the decays.     
   
\vskip 0.5cm
\end{abstract}
%
%

PACS Numbers: 11.30.Er, 11.80.Et, 13.25.-k, 13.30.Eg 
%
%
%
\end{titlepage}
\newpage
\noindent

\section{Introduction}

The physics of the $B$-meson has opened a new door in the sector of hadronic weak interactions. 
In particular, numerous $B$ decays, analyzed during more than a decade, have confirmed the CKM mechanism 
for CP violation\cite{bs}; yet, it is generally believed that some particular decays of this kind could 
reveal new physics beyond the Standard Model (SM). Therefore several experiments in this sense have been 
suggested or even realized. In particular, data of $B^{\pm} \to K^{\pm} p {\bar p}$ decay have been 
published recently by the LHCb collaboration[2-4].

Three-body baryon-antibaryon $B$ decays were detected for the first time by ARGUS in 1987\cite{ar0}. Although immediately ruled out by CLEO\cite{cl0}, that claim gave 
rise to a lively theoretical interest in the subject for several years: see ref. \cite{chg} for a complete and thorough review. More 
recently, this interest was revived[8-12] and new experimental results were found by the
CLEO\cite{cl}, Belle\cite{be1,be2} and BaBar\cite{bb1,bb2} collaborations.

A three-body decay presents some advantages over a two-body one. For example, if one chooses a three-body charged $B$ decay such that all final particles are charged, its detection is favored. Moreover, an amplitude analysis in different intervals of the Dalitz plot allows direct measurement of the relative strong phases\cite{ltm}.

Baryonic three-body decays of charged $B$ are especially useful for studying the strong dynamics and, in principle, for analyzing 
the observables sensitive to new physics\cite{hs}. Indeed, it is customary to assume that final state interactions\cite{mrr} (FSI) are much smaller\cite{lhcb} than those of three-meson decays, characterized by large re-scattering effects\cite{mrr,wf,ltm}. This fact would allow the study of FSI by comparison between the two different kinds of decays\cite{lhcb} and therefore could help in determining the weak phase, which generally presents serious difficulties\cite{ads}. However, as we shall see in the present paper, FSI might play a surprisingly important role also in baryonic decays.
 
Besides, such decays are somewhat advantageous with respect to the baryonic two-body ones (e. g., $B^0 \to p \bar{p}$), 
in that they give rise to greater decay rates than two-body baryonic 
decays. This is connected with the threshold baryon-antibaryon peaks,  predicted by Hou and Soni\cite{hs} as a consequence 
of the increasing difficulty of a quark to hadronize to a baryon at higher energies. In particular, if the final state includes the 
proton-antiproton system, the effect is attributed to 
re-scattering\cite{hms}. Moreover, the decays considered 
exhibit intriguing forward-backward (FB) and Dalitz plot asymmetries\cite{bb2,be2,lhcb}.

This kind of decays is generally described by assuming two factorizable amplitudes of the type\cite{cht,chg}
\begin{equation} 
{\cal T} = \langle M|(\bar{q}_3 q_2)|0\rangle \langle {\cal B}_1{\bar{\cal B}}_2|(\bar{q_1} b)|B^-\rangle \ ~~~~ \  \mathrm{and} 
\ ~~~~ \ {\cal I} = \langle {\cal B}_1{\bar{\cal B}}_2|(\bar{q}_1 q_2)|0\rangle 
\langle M|(\bar{q_3} b)|B^-\rangle, \nonumber
\end{equation}
called respectively transition and current terms. Here $M$, ${\cal B}_1$ and ${\bar{\cal B}}_2$ are the final meson, baryon and antibaryon respectively, while $b$ and $q$ ($\bar{q}$) are the destruction (creation) operators of the active (light) quarks involved in the decay.  

Turning to the decay $B^{\pm} \to K^{\pm} p \bar{p}$ (to be named $K$ decay in the following), it presents two main differences with respect to the  $B^{\pm} \to \pi^{\pm} p \bar{p}$ ($\pi$ decay from now on). Firstly, it has a wider $p\bar{p}$ peak, which may be explained as an effect of the interference between the transition and the current 
contributions\cite{cht}: indeed, while suppressed in the $\pi$ decay, the current term contributes considerably to the $K$ decay and decreases more slowly than the transition term at increasing 
$p\bar{p}$ effective mass. Secondly, the sign of the FB asymmetry ($A_{FB}$) in the $\pi$ decay is opposite to that in 
the $K$ decay: while the behavior of the $\pi$ asymmetry can be explained in the framework of quark dynamics\cite{chg}, the other one 
is more difficult to interpret.

An important feature of the $K$ decay is that it derives contributions from two different amplitudes, typically tree and penguin, endowed with different weak and strong phases. Therefore, we expect that data exhibit a sizable direct CP asymmetry ($A_{CP}$), to be compared with the SM predictions. Indeed, this observable has been measured recently to high precision by the LHCb collaboration\cite{lhcb}. Unfortunately, as recalled above, comparison with theory is particularly difficult in weak decays, owing to the problem of disentangling real CP-violation effects from strong FSI. Therefore, in the interest of understanding and parametrizing as precisely as possible the effects of non-perturbative QCD in hadronic weak decays, we perform a best fit to some available data, like the differential branching fraction and the asymmetries $A_{FB}$ and $A_{CP}$. More precisely, on the one hand,
we adopt for the decay amplitude the parametrization by ref. \cite{cht}; on the other hand, however, we modify some parameters by means of a best fit, in view of various considerations. As we shall see, our analysis, based on the data by Belle\cite{be2} and on the most recent data of LHCb\cite{lhcb}, leads to conclusions in contrast with previous ones; in particular, we give a different answer to one of the main questions illustrated above, that is, the origin of the FB asymmetry. Lastly, we make some predictions about other observables measurable in principle. 
   
Section 2 is devoted to the definition of some observables, either presently available or measurable in the future, and to the theoretical function used for interpreting them. In sect. 3, we present our phenomenological analysis, both assuming factorization and relaxing this assumption; moreover we show some predictions of our model. Lastly, in sect. 4, we discuss the results obtained and draw some conclusions.

\section{Observables and Theoretical Function}

\subsection{Observables}

We fit the data relative to the the differential branching fraction of the decay $B^{\pm} \to K^{\pm} p \bar{p}$\cite{lhcb,be2}, i. e.,
\begin{equation} 
\gamma =  \frac{dBf_{Kp{\bar p}}}{dm_{p{\bar p}}}. \label{dgm}
\end{equation}
The LHCb data\cite{lhcb} are re-scaled in such a way that the integral over all of the $p \bar{p} K$ spectrum equals the total branching fraction\cite{pdg}, i. e.,
\begin{equation} 
Bf(B^{\pm} \to K^{\pm} p \bar{p}) = (5.9 \pm 0.5)\cdot 10^{-6}. \label{ebf}
\end{equation}

On the other hand, we calculate the CP asymmetry and the FB asymmetry, whose
experimental values are, respectively\cite{lhcb},
\begin{equation} 
 A_{CP} = -0.022 \pm 0.031 \pm 0.007 \label{bf-}
\end{equation}
and 
 \begin{equation} 
  A_{FB} = 0.370 \pm 0.018\pm 0.016. \label{dbf}
 \end{equation}
The latter is defined as
\begin{equation} 
  A_{FB} = \frac{N^+-N^-}{N^++N^-}
 \end{equation}
and $N^{\pm}$ is the number of events for which cos$\theta_p$ is
positive (negative), $\theta_p$ being the angle between the meson and the opposite-sign baryon in the $p\bar{p}$ rest frame. This definition holds as well for the $\pi$ decay.

\subsection{Theoretical Function}

\subsubsection{Matrix Element of the Decay}

We parametrize the decay amplitude for $B^{\pm} \to K^{\pm} p \bar{p}$, according to Chua {\it et al.}\cite{cht} (see also refs. \cite{gg1,chg}),
i. e.,
\begin{equation}
{\cal M} = \frac{G_F}{\sqrt{2}}({\cal I}+ {\cal T});
\end{equation}
here $G_F$ is, as usual, the Fermi constant of weak interactions and  ${\cal I}$ and ${\cal T}$ are respectively the current and transition
terms, according to the definition given in the introduction.
As regards the current term, we have
\begin{eqnarray}
{\cal I} &=& {\cal I}_1 + {\cal I}_2,
\\
{\cal I}_1 &=& \langle K^-|L_{sb}^{\mu}|B^-\rangle (\sum_{i=1}^4\alpha_i
\langle p \bar{p}|L_{\mu}^i|0\rangle + \alpha_5 \langle p \bar{p}|R_{\mu}|0\rangle),
\label{I1}
\\
{\cal I}_2 &=& \langle |K^-|\tilde{L}_{sb}|B^-\rangle \alpha_6 \langle p 
\bar{p}|\tilde{R}_{ss}|0\rangle.
\label{I2}
\end{eqnarray}
On the other hand, the transition term reads as
\begin{equation}
{\cal T} = \langle K^-|\tilde{L}_{us}|0\rangle (\alpha_7\langle p 
\bar{p}|S_{ub}| B^-\rangle + \alpha_8\langle p \bar{p}|P_{ub}|B^-\rangle). \label{trt}
\end{equation}
Here we have set
\begin{eqnarray}
L_{sb}^{\mu} &=& \bar{s}_L\gamma^{\mu} b_L, \ ~~~~ \ L^1_{\mu} = 
\bar{u}_L \gamma_{\mu} u_L, \ ~~~~ \ L^2_{\mu} = \bar{u}_L\gamma_{\mu} u_L + 
\bar{d}_L\gamma_{\mu} d_L + \bar{s}_L\gamma_{\mu} s_L, 
\\
L^3_{\mu} &=& \bar{s}_L\gamma_{\mu} s_L, \ ~~~~ \ L^4_{\mu} = e_u\bar{u}_L\gamma_{\mu} u_L + 
e_d \bar{d}_L\gamma_{\mu} d_L + e_s\bar{s}_L\gamma_{\mu} s_L,  
\\
R_{\mu}&=& \bar{u}_R\gamma_{\mu} u_R + \bar{d}_R\gamma_{\mu} d_R + \bar{s}_R\gamma_{\mu} s_R,
\ ~~~~ \  \tilde{L}_{sb} = \bar{s}_L b_L, 
\\ 
\tilde{R}_{ss} &=& \bar{s}_R s_R, \ ~~~~ \ \tilde{L}_{us} = \bar{s}_L u_L, \ ~~~~ \    
S_{ub} = \bar{u} b, \ ~~~~ \ P_{ub} = \bar{u} \gamma_5 b.  
\end{eqnarray}
Moreover the $\alpha$ are connected to the CKM matrix elements and to the Wilson coefficients. Focusing on the $B^-$ decay, we have 
\begin{eqnarray}
\alpha_1 &=& V_{ub} V^*_{us} a_2, \ ~~~~ \ \alpha_2 = -V_{tb} V^*_{ts} a_3, \ ~~~~ \ \alpha_3 =
-V_{tb} V^*_{ts} a_4, 
\\
\alpha_4 &=& -\frac{3}{2} V_{tb} V^*_{ts} a_9, \ ~~~~ \ \alpha_5 = -V_{tb} V^*_{ts} a_5, \ ~~~~ \ 
\alpha_6 = 2V_{tb} V^*_{ts} a_6,
\\
\alpha_{7,8} &=& V_{ub} V^*_{us} a_1-V_{tb} V^*_{ts} a_{\pm}, \ ~~~~ \ a_{\pm} = a_4 \pm 
a_6 \frac{2 m^2_K}{m_b(m_s+m_u)}.   
\end{eqnarray}
The numerical values of the $a$ and of the $V$ are listed in the Appendix. The formulae of the $B^+$ decay are obtained by taking the complex conjugates of the CKM matrix elements. Since $V_{ub}$ is the only complex CKM matrix element in the equations above, only $\alpha_1$, $\alpha_7$ and $\alpha_8$ are involved in the CP asymmetry. 

The various (non-perturbative) 
matrix elements of the quark operators are connected to some form factors, as shown in the Appendix. As a result we get  
\begin{equation}
{\cal M} = \frac{G_F}{\sqrt{2}} \frac{1}{2m_p}\sum_{l=1}^4 \beta_l \bar{u}(p_p)O_l v(p_{\bar p}). 
\label{mel}
\end{equation}
Here $p_p$ and $p_{\bar p}$ are the four-momenta of the proton and of the antiproton respectively;
$u$ and $v$ are their standard Dirac spinors, normalized as $\bar{u} u$ = $-\bar{v} v$ = $2m_p$, where $m_p$ is the proton mass. 
Moreover, 
\begin{equation}
O_1 = p\hspace{-0.45 em}/_B, \ ~~~~ \ O_2 = p\hspace{-0.45 em}/_B \gamma_5, \ ~~~~ \ O_3 = I, 
\ ~~~~ \ O_4 =  \gamma_5,
\end{equation}
with $p_B$ being the four-momentum of the decaying resonance $B$. Lastly, the coefficients $\beta$ are
\begin{eqnarray}
\beta_1 &=& 2F_1\sum_{i=1}^4 \Phi^i\xi_i - f_K m_b \alpha_8 F_{V5}, \label{bt1}
 \ ~~~~ \ \ ~~~~ \  \ ~~~~ \ \ ~~~~ \ \ ~~~~ \  \ ~~~~ \ 
\\
\beta_2 &=& 2F_1\sum_{i=1}^4 g_A^i \eta_i - f_K m_b \alpha_7 F_A,
\ ~~~~ \ \ ~~~~ \  \ ~~~~ \ \ ~~~~ \ \ ~~~~ \  \ ~~~~ \ 
\\
\beta_3 &=& \frac{\ell\cdot\delta}{2m_p} F_1\sum_{i=1}^4 \xi_i\hat{\Phi}^i - f_K m_b \alpha_8 F_P, \label{bt3}
\ ~~~~ \ \ ~~~~ \  \ ~~~~ \ \ ~~~~ \ 
\\
\beta_4 &=& F_0 [\frac{r p_K^2}{2m_p}\sum_{i=1}^4\eta_i h_A^i + \alpha_6\frac{m_B^2-m_K^2}{m_b-m_s}
(\frac{m_p}{m_s}g_A^3+\frac{p_K^2}{4m_p m_s}h_A^3)] + \nonumber 
\\
&-& f_K m_b \alpha_7 F_P - 2(r+1)m_p F_1\sum_{i=1}^4 g_A^i \eta_i + 2m_p (r F_0 \sum_{i=1}^4 g_A^i \eta_i 
+ f_K m_b \alpha_7 F_A). \label{bt4}
\end{eqnarray}
Here
\begin{eqnarray}
\xi_i(\eta_i) = \pm\alpha_i+\alpha_5\delta_{i,2}, ~~ (i = 1, 2, 3, 4), \ ~~~~ \ \ ~~~~ \
r = \frac{m_B^2-m_K^2}{(p_{\bar p} + p_p)^2}, 
\\
l = 2p_B -(1+r)(p_{\bar p} + p_p),  \ ~~~~ \ \ ~~~~ \ \ ~~~~ \ \ ~~~~ \ \delta = p_{\bar p} - p_p,  
\end{eqnarray}
$p_K$ is the four-momentum of the $K$-meson, $m_{b(s)}$ the masses of the quarks and $m_K$ and $m_B$ the meson masses. Moreover $f_K$ is the decay constant of the $K$-meson and $F_0$, $F_1$, $F_A$, $F_P$, $F_{V5}$, $\Phi_i$, $\hat{\Phi}_i$, $g_A^i$ and  $h_A^i$ ($i$ = 1 to 4) are form factors. In particular, the $\hat{\Phi}_i$ - related to the time-like nucleon form factors, as well as the $\Phi_i$ - have been set equal to zero, according to ref. \cite{cht}; we shall discuss this assumption in sect. 4. In this connection, it is worth noting that $F_{0,1}$, as well as the corresponding proportionality factors $S_{0,1}$, defined in the Appendix, have the dimensions of an energy, unlike assumed in refs. \cite{cht,mvs}.

Lastly, as we shall see in the next section, our analysis involves also the $\pi$ decay, for which some variations have to be performed: firstly, one has to replace $s_{L(R)}$ with $d_{L(R)}$, $f_K$ with $f_{\pi}$, $m_s$ with $m_d$ and $V_{us}$, $V_{ts}$ with $V_{ud}$, $V_{td}$; secondly, the values of some $a_i$ and parameters in the form factors have to be changed, as shown in the Appendix. 

\subsubsection{Calculations of Observables}
The modulus squared of the matrix element (\ref{mel}) reads as
\begin{equation}
|{\cal M}|^2 = \frac{G_F^2}{2} \frac{1}{4m_p^2} \{\sum_{l=1}^4 |\beta_l|^2\Omega_{ll} + 2
[\Omega_{13}\Re(\beta_1\beta_3^*) + \Omega_{24}\Im(\beta_2\beta_4^*)]\}.  \label{sq}
\end{equation}
Here 
\begin{eqnarray}
\Omega_{11} &=& 4[2p_B\cdot p_p p_B\cdot p_{\bar{p}} - m_B^2(m_p^2+p_p\cdot p_{\bar{p}})],
\\
\Omega_{22} &=& 4[2p_B\cdot p_p p_B\cdot p_{\bar{p}} + m_B^2(m_p^2-p_p\cdot p_{\bar{p}})],
\\
\Omega_{33} &=& 4(-m_p^2+p_p\cdot p_{\bar{p}}), \ ~~~~ \ \ ~~~~ \ \ ~~~~ \  \ ~~~~ \ 
\\
\Omega_{44} &=& 4(m_p^2+p_p\cdot p_{\bar{p}}), \ ~~~~ \ \ ~~~~ \ \ ~~~~ \  
\ ~~~~ \ 
\\
\Omega_{13} &=& 4m_p p_B\cdot (p_{\bar{p}}-p_p), \ ~~~~ \ \ ~~~~ \ \ ~~~~ \  \ ~~~~ \ \label{ifl}
\\
\Omega_{24} &=& 4m_p p_B\cdot (p_{\bar{p}}+p_p). \ ~~~~ \ \ ~~~~ \ \ ~~~~ \  \ ~~~~ \ 
\end{eqnarray}

Then the differential decay width is 
\begin{equation}
d \Gamma = \frac{(2\pi)^4}{2m_B}|{\cal M}|^2 \Pi_{i=1}^3
\frac{4m_p^2 d^3p_i}{(2\pi)^32E_i}
\delta(m_B-\sum_{i=1}^3E_i)\delta^3(\sum_{i=1}^3{\bf p}_i), \label{ddw0}
\end{equation}
where $E_i$ and ${\bf p}_i$ ($i$ = 1 to 3) are the energies and the momenta, respectively, of  the proton, 
of the antiproton and of the $K$-meson in the $B$ rest frame. By integrating over all variables but the 
momentum of the proton, we get
\begin{equation}
d \Gamma = \frac{m_p^2 }{4(2\pi)^4 m_B} |{\cal M}|^2 p_3\frac{d^3p_1}{E_1 E_2},
\end{equation}
where $p_3$ = $|{\bf p}_3|$. In order to calculate the FB (or helicity\cite{lhcb}) asymmetry, it is convenient
to express energies and momenta as functions of kinematic quantities in the $p\bar{p}$ rest frame. To this end, we perform a Lorentz boost from that frame to the $B$ rest frame. Moreover we integrate over the azimuthal angle of the proton: since $|{\cal M}|^2$ does not depend on it, we get
\begin{equation}
d \Gamma = \frac{J m_p^2  p_3 \pi_p^2}{2^4 \pi^3 m_B} |{\cal M}|^2 (\frac{1}{4}E_0^2-\frac{p_3^2}{t}\pi_z^2)^{-1}
d\pi_p d cos\theta_p.
\end{equation}
Here $\pi_z$ = $\pi_p cos\theta_p$, $\pi_p$ is the modulus of the momentum of the proton and of the antiproton in the $p\bar{p}$ frame and $\theta_p$ the helicity angle\cite{lhcb}, between the meson and the
opposite-sign baryon in the same frame; moreover $t$ = $4(m_p^2+\pi_p^2)$ is the effective mass squared of the $p\bar{p}$ system, $E_0$ = $\sqrt{t+p_3^2}$ and  
\begin{equation}
 J = \frac{E_0}{\sqrt{t}}-4\pi_z\frac{dp_3}{dt}+\frac{8\pi_z^2}{\sqrt{t}}[\frac{1}{2E_0}(1+\frac{dp_3^2}{dt})
 -\frac{E_0}{2t}].\label{jcb}
\end{equation}
Taking into account energy-momentum conservation, we have
\begin{equation}
p_3 = \frac{1}{2m_B}[(m_B^2-m_K^2)^2-2t(m_B^2-m_K^2)+t^2]^{1/2}.
\end{equation}
Our analysis requires the formulae for the differential decay width and for the differential FB difference, as functions of the $p\bar{p}$ invariant mass $m_{p\bar{p}}$ = $\sqrt{t}$. To this end, we perform the appropriate integrations over cos$\theta_p$ - respectively $\int_{-1}^1$ and $\int_0^1-\int_{-1}^0$ - expressing the Lorentz invariant coefficients $\Omega_{ij}$ in terms of the kinematic variables just defined. The results are
\begin{equation}
\frac{d \Gamma}{dm_{p\bar{p}}}(\frac{d \Delta\Gamma}{dm_{p\bar{p}}}) = \frac{G_F^2}{2} \frac{m_{p\bar{p}} \pi_p p_3}{4^4\pi^3
m_B} I(\pi_p) [\Delta I(\pi_p)]. \label{ddw}
\end{equation}
Here 
\begin{equation}
I(\pi_p) = \sum_{i=1}^3 \rho_i G_i, \ ~~~~ \ \ ~~~~ \ \Delta I(\pi_p) = \sum_{i=1}^3 \rho_i \Delta G_i,
\end{equation}
with
\begin{eqnarray}
\rho_1 &=& 8m_B^2 (|\beta_1|^2+|\beta_2|^2), \ ~~~~ \ \ ~~~~ \ \ ~~~~ \ \ ~~~~ \ 
\\
\rho_2 &=& -8\{m_B^2[(m_p^2+\pi_p^2)|\beta_1|^2+\pi^2_p(|\beta_2|^2]-\pi^2_p(|\beta_3|^2+ \nonumber
\\
&+&[(m_p^2+\pi_p^2)|\beta_4|^2-m_p m_B E_0\Im(\beta_2\beta_4^*)\}, \ ~~~~ \ ~~~ \ ~~~ \ 
\\
\rho_3 &=& 16 m_p m_B p_3 t^{-1/2}\Re(\beta_1\beta_3^*), \ ~~~~ \ \ ~~~~ \ \ ~~~~ \ 
\end{eqnarray}

\begin{eqnarray}
G_1 &=& 2(A+\frac{1}{3}C), \ ~~~~ \ 
G_2 = \frac{1}{h} [(\frac{A}{a}+aC)~ln\frac{a+1}{|a-1|}-2C]  
\\
G_3 &=& \frac{\pi_pB}{h}(2-a ~ ln\frac{a+1}{|a-1|}), \ ~~~~ \ \ ~~~~ \ 
\ ~~~~ \ \ ~~~~ \ \ ~~~~ \ \ ~~~~ \
\\
\Delta G_1 &=& -B, \ ~~~~ \ \Delta G_2 = \frac{B}{h} ln\frac{|a^2-1|}{a^2},  \ ~~~~ \ \ ~~~~ \ 
\\
\Delta G_3 &=& -\frac{\pi_p}{h} [(A+Ca^2)ln\frac{|a^2-1|}{a^2}+C] 
\ ~~~~ \ \ ~~~~ \ \ ~~~~ \ \ ~~~~ \ 
\end{eqnarray}
and
\begin{eqnarray}
A &=& \frac{E_0}{\sqrt{t}}, \ ~~~~ \  B = 4\pi_p \frac{dp_3}{dt}, 
\ ~~~~ \ C = \frac{8\pi_p^2}{\sqrt{t}}[\frac{1}{2E_0}(1+\frac{dp_3^2}{dt})-\frac{E_0}{2t}], 
\\
a &=& \frac{E_0\sqrt{t}}{2\pi_p p_3}, \ ~~~~ \ \ ~~~ \ \ ~~~ \ h = \frac{p_3^2\pi_p^2}{t}. \ ~~~~ \ ~~~~ \  
\end{eqnarray}
Eqs. (\ref{ddw}) allow one to calculate several observables, to be compared with data. In particular, we are interested in the quantities defined in subsect. 2.1 - that is, the differential and total branching fractions 
and the overall FB and CP asymmetries - and in the differential CP asymmetry, 
\begin{equation}
{\cal A}_{CP} = \frac{\gamma^- - \gamma^+}{\gamma^- + \gamma^+},
\label{dcp}
\end{equation}
where $\gamma$ is defined by eq. (\ref{dgm}).

\section{Phenomenological Analysis}

Here we compare the theoretical function just written with the available observables listed in subsect. 2.1, that is, the differential branching fraction and the overall CP and FB asymmetries. In particular, as regards the first observable, we exclude the contribution of the two charmonium bands, around 3.1 GeV and around 3.6 GeV. Concerning the form factors involved in eq. (\ref{ddw}), firstly we adopt the factorization assumption, like Chua {\it et al.}\cite{cht}. However, as we shall see, this disagrees with data\cite{lhcb,be2}. Therefore we shall propose a modification of the model, in view of some considerations about the 
main contributing graphs.

\subsection{Factorization Assumption} 

As a first attempt, we assume factorization for the current and transition terms, according to refs. \cite{chg,cht,gg1,gg2} and refs. therein. Indeed, this scheme generally describes the three-body baryonic $B$ decays in a satisfactory way; in particular, it seemed to be supported\cite{cht} by the early data of the $K$ and $\pi$ decays\cite{be3}. Therefore a comparison with the most recent data of those decays\cite{lhcb,be2} is in order.
 
In the context of factorization, the transition term has the same proportionality coefficient for the $K$ decay as for the $\pi$ decay; 
as regards the current form factors, the proportionality coefficients have been calculated by Melikhov and Stech\cite{mvs} (see also ref. \cite{cht}). 

In the case of the $\pi$ decay, the penguin amplitude is subdominant with respect to the color-allowed tree diagram\cite{bb2}, even taking into account re-scattering effects to be discussed in subsect. 3.2. Therefore, the current term, which derives contributions almost exclusively from the penguin amplitude, is suppressed with respect to the transition term, which consists, instead, of tree and penguin contributions. Incidentally, in that decay, also a subdominant, color-suppressed tree diagram has to be accounted for. It corresponds, in the case of the $B^-$ decay, to the quark subprocess $b \to u\bar{u}d$, followed by recombination of the $d$-quark with the spectator $\bar{u}$ to form the $\pi^-$, and by the fragmentation of the active $u\bar{u}$ pair into $p\bar{p}$; the $\bar{p}$ is next to the $\pi^-$ in rapidity space and can resonate with it as a $\Delta^{--}$, giving a negative contribution to the FB asymmetry. 
  
The above evaluation about the dominating diagrams agrees qualitatively with the numerical results of the current form factors given in ref. \cite{mvs}; indeed, the current term turns out to be negligibly small for the $\pi$ decay\cite{cht}. Therefore, in a simplifying assumption, we neglect the current term; we impose as well the functions $F_{V5}$ and $F_P$ to vanish, according to ref. \cite{cht}. Then, our expression of the branching fraction depends only on $F_A$, whose proportionality coefficient is $C_A$; imposing the branching fraction to equal the experimental value\cite{pdg}, 
\begin{equation}
Bf(B^{\pm} \to \pi^{\pm} p \bar{p}) = (1.62\pm 0.20)\cdot 10^{-6},
\end{equation}
yields
\begin{equation}
|C_A| = (42.1^{+2.4}_{-2.6}) ~ GeV^5. \label{trs}
\end{equation}
Factorization implies that this contribution - up to the sign of $C_A$, and up to some constants, as explained at par. 2.2.2 - should be present also in the $K$ decay. Here, however, also the current term is important\cite{cht}, since, in this case, it receives contributions exclusively from the penguin amplitude, which is dominant over the tree amplitude. This fact is confirmed by the sizes of the proportionality coefficients of the current form factors for that decay\cite{mvs}. 
Therefore, in order to calculate the current term, we insert into eqs. (\ref{I1}) and  (\ref{I2}) the formulae of the electromagnetic nucleon form factors in the time-like region\cite{cht,mvs}, shown diagrammatically in fig. \ref{fig:gamma}; such formulae are given in the Appendix, while the proportionality coefficients of the form factors, $S_0$ and $S_1$, are listed in Table \ref{tab:one}\cite{cht}.
\begin{figure}[h]
\centering
\includegraphics[width=0.40\textwidth] {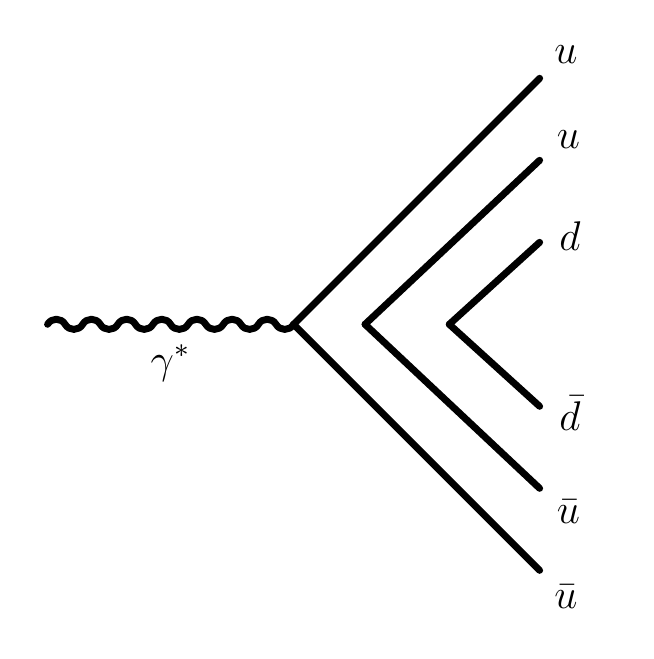}
\caption{Diagram of annihilation of a virtual $\gamma$ into $ p \bar p$ \label{fig:gamma}} 
\end{figure}

As regards the transition term, we assume the value (\ref{trs}) for 
$C_A$, with the two possible signs. The choice of the - sign yields
\begin{equation}
Bf(B^{\pm} \to K^{\pm} p \bar{p}) = (2 .96 \pm 0.27)\cdot 10^{-6},
\end{equation}
which is considerably smaller than the experimental value (\ref{ebf}). The choice of the + sign for $C_A$ yields even a lower value. Also data\cite{lhcb,be2} of the differential branching fraction are in disagreement with the factorization assumption, as shown by the dashed line in fig. \ref{fig:experiment}; therefore  we conclude that this assumption is inadequate. The situation is somewhat analogous to the one illustrated in ref. \cite{ccs} about the decays of $B$ to $K \chi_0$, to $D^{(*)0} \pi^0$ and to $\pi^0 \pi^0$, which are surprisingly enhanced with respect to the factorization assumption and demand both 
non-factorizing terms and re-scattering effects.  

\subsection{Remarks on the Main Contributing Graphs}

\begin{figure}[h]
\centering
\includegraphics[width=0.95\textwidth] {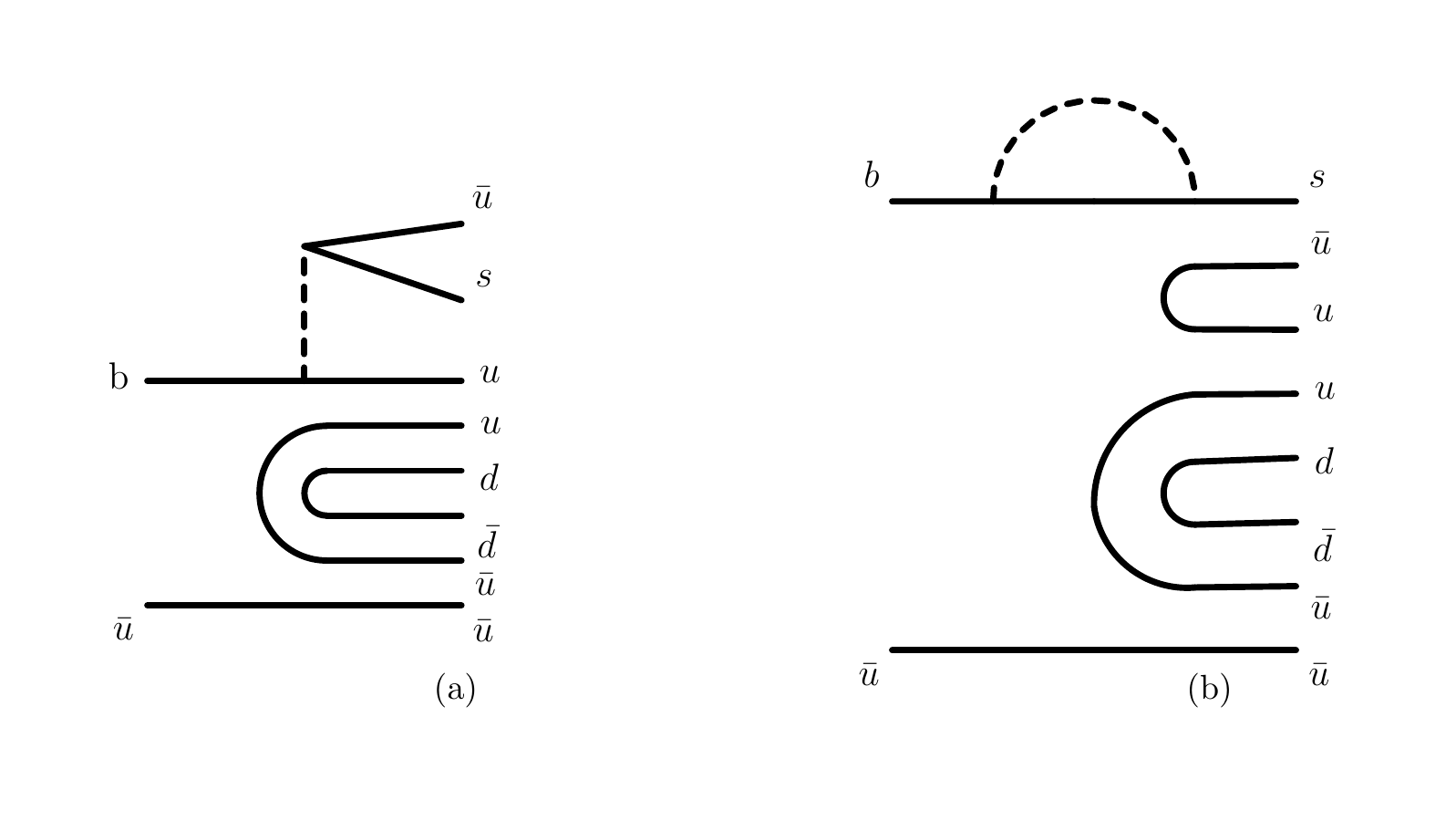}
\caption{Main diagrams in the decays $B^{\pm} \to K^{\pm} p \bar{p}$ \label{fig:feyn}}
\end{figure}

After the result just shown above, a criticism of the assumptions
made is in order. Therefore we analyze in detail
the main graphs contributing to the decay considered, 
which we illustrate in fig. \ref{fig:feyn}. 
\vskip 1cm

Figs. 2a and 2b represent, respectively, the color-allowed and the penguin diagram. According to the former one, the meson is by no means correlated to the baryons, since the quark hadronization to the meson occurs just after the weak (very short-ranged) interaction; therefore factorization could be of course assumed for this diagram, which contributes to the transition term. But as told, in the $K$ decay, this amplitude is sub-dominant with respect to the penguin one. This latter consists of several components contributing to the 
$K$ and/or $\pi$ decay, described and represented in detail in refs. \cite{chyn} and\cite{chg}: some of them may be regarded as factorizable, but not all. The evident failure of factorization leads us to conclude that non-factorizing diagrams play quite an important role in the $K$ decay. In particular, we point out a contribution fed by the quark subprocess $b\to c\bar{c}s$. It may give rise to a $p\bar{p}$ pair through re-scattering from $\Lambda_c \bar{\Lambda}_c$ pair formation, which can occur either by $c\bar{c}$ fragmentation, or via intermediate states like $D^{(*)} \bar{D}_s^{(*)}$, $\Xi_c\bar{\Lambda}_c$\cite{chg}; this causes an anomalously large enhancement of the decay rate with respect to the naive factorization\cite{chg}. In the $\pi$ decay, the
corresponding amplitude is suppressed by a factor $|V_{cd}/ V_{cs}|$ with
respect to the $K$ decay. Since the current term interferes only weakly with the transition term\cite{cht}, this process gives a negligibly small contribution to the differential decay width, as anticipated in subsect. 3.1. On the contrary, it might explain the prevalent role of the current term in the $K$ decay, as we shall establish in subsect. 3.3.

The failure of factorization is also confirmed by the numerical results presented by Chua {\it et al.}\cite{cht}, 
since the current contribution depends crucially on the effective number of colors, $N_c$ (see also ref. \cite{gg2}). On the contrary, 
the fit to the $\pi$ decay, dominated by the tree diagram, is stable versus $N_c$; this is reflected also on the transition term of the 
$K$ decay\cite{cht}, for which those authors assume the same parametrization as for $\pi$. The fit to $B^{\pm} \to \pi^{\pm} \Lambda \bar{p}$, which depends essentially on a factorizable penguin diagram\cite{chyn}, is also reasonably stable versus $N_c$.

Aside from that, as told, the long-range FSI might be not so negligible, somewhat analogously to the case of $B$ decays to mesons\cite{ccs}, although on a smaller scale. This may contribute to the failure of factorization and may give rise to further effects, to be discussed in sect. 4.

\subsection{An Alternative Parametrization}

\begin{figure}[h]
\begin{center}
\includegraphics[width=0.70\textwidth] {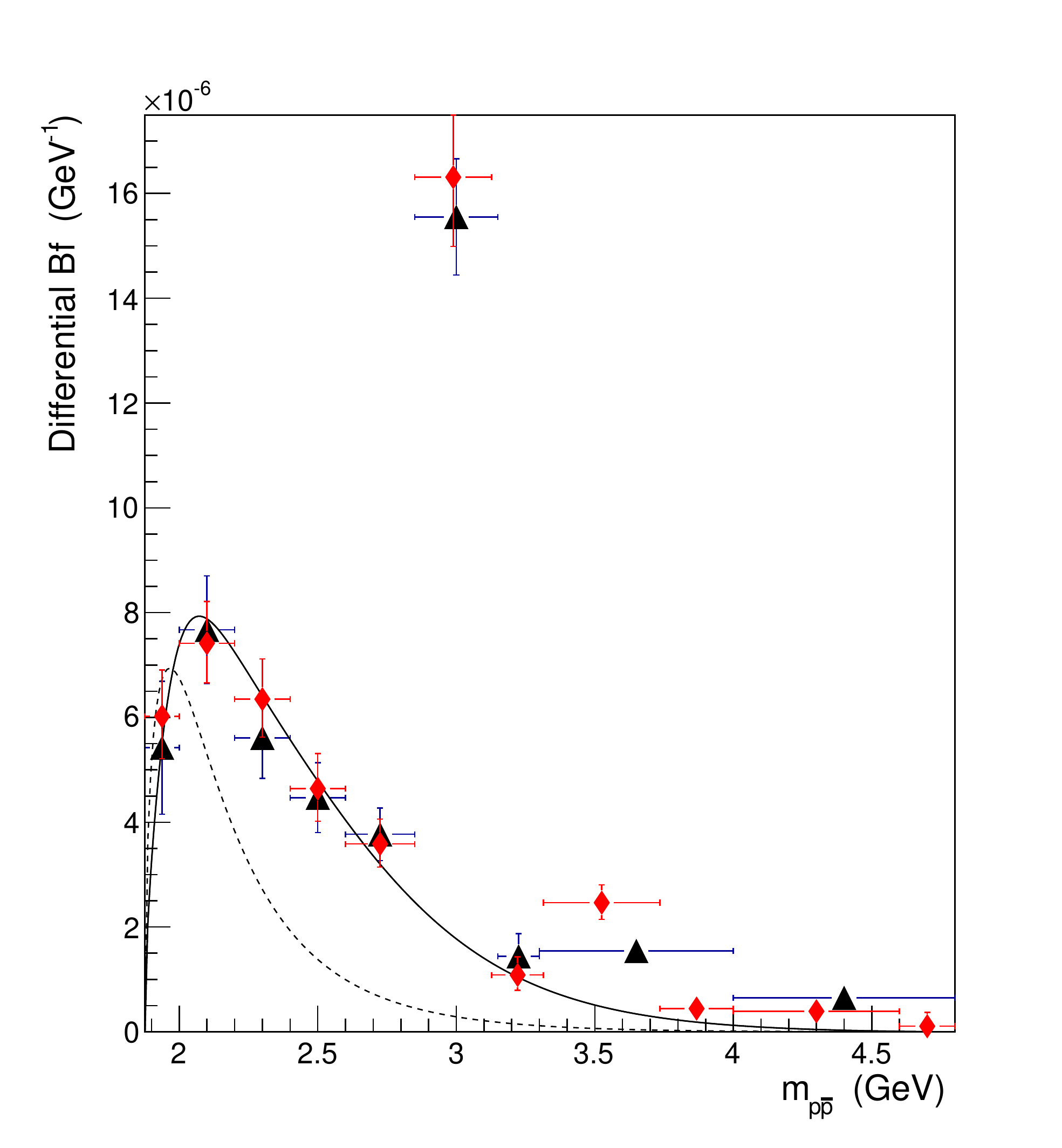}
\caption{$B^{\pm}\to K^{\pm}p\bar{p}$ decay: differential branching fraction {\it vs} $m_{p\bar p}$, charge averaged case. 
LHCb and Belle data are 
represented, respectively, by rhombuses and triangles. The continuous (dashed) line refers to the best fit (factorization assumption).}
\label{fig:experiment}
\end{center}
\end{figure}

All of the above considerations suggest to relax the factorization assumption. Indeed, we have to take into account both non-factorizable
terms and FSI; these latter are typically non-perturbative and therefore 
cannot be deduced from QCD principles. To this end, we regard $C_A$, $C_{V5}$, $C_P$, $S_0$ and $S_1$ as free parameters, which we determine by minimizing the $\chi^2$ in a fit to data, that is, as told, the differential branching fraction\cite{lhcb,be2} and the asymmetries $A_{FB}$ and $A_{CP}$\cite{lhcb}. In order to avoid strong correlations between the various parameters, we set $C_{V5}$ = $C_P$ = 0, like Chua {\it et al.}\cite{cht}. The parameters of the fit are shown in Table \ref{tab:one}, where they are compared to those of factorization. Furthermore the fit to the $m_{p{\bar p}}$ 
distribution is exhibited in fig. \ref{fig:experiment}, continuous line. 
 
\begin{table*}[h!]
\setlength{\tabcolsep}{1.5pc}
\newlength{\digitwidth} \settowidth{\digitwidth}{\rm 0}
\catcode`?=\active \def?{\kern\digitwidth}
\caption{Parameters of the theoretical function. F1 refers to factorization, F2 shows the results of the best fit. 
$C_A$ and $C_{V5}$ are expressed in $GeV^5$, $C_P$ in $GeV^8$, $S_0$ and $S_1$ in $GeV$.} 
\label{tab:one}
 \begin{tabular*}{\textwidth}{l r c c r r }
\hline
Fit    &  $C_A~~~~$    &    $~~C_{V5}$    &    $~~C_P$   &   $S_0~~~~ $   &    $S_1~~~~$   \\
\hline
F1    & $-42.1\ ^{+2.4}_{-2.6}$  &        --       &    --     &       0.36~~~                      &     0.36~~~               \\[5pt]
F2    &  $-66.0^{+24.8}_{-\ 9.1}$  &      --      &    --     &  $ ~~187.3^{+\ 48.2}_{-126.9} $      &     $-1.25^{+0.05}_{-0.03} $ \\[5pt]
\hline
\end{tabular*}
\end{table*}

\vskip 1cm




\subsection{Results and Model Predictions}
\begin{figure}[h]
\begin{center}
\includegraphics[width=0.70\textwidth] {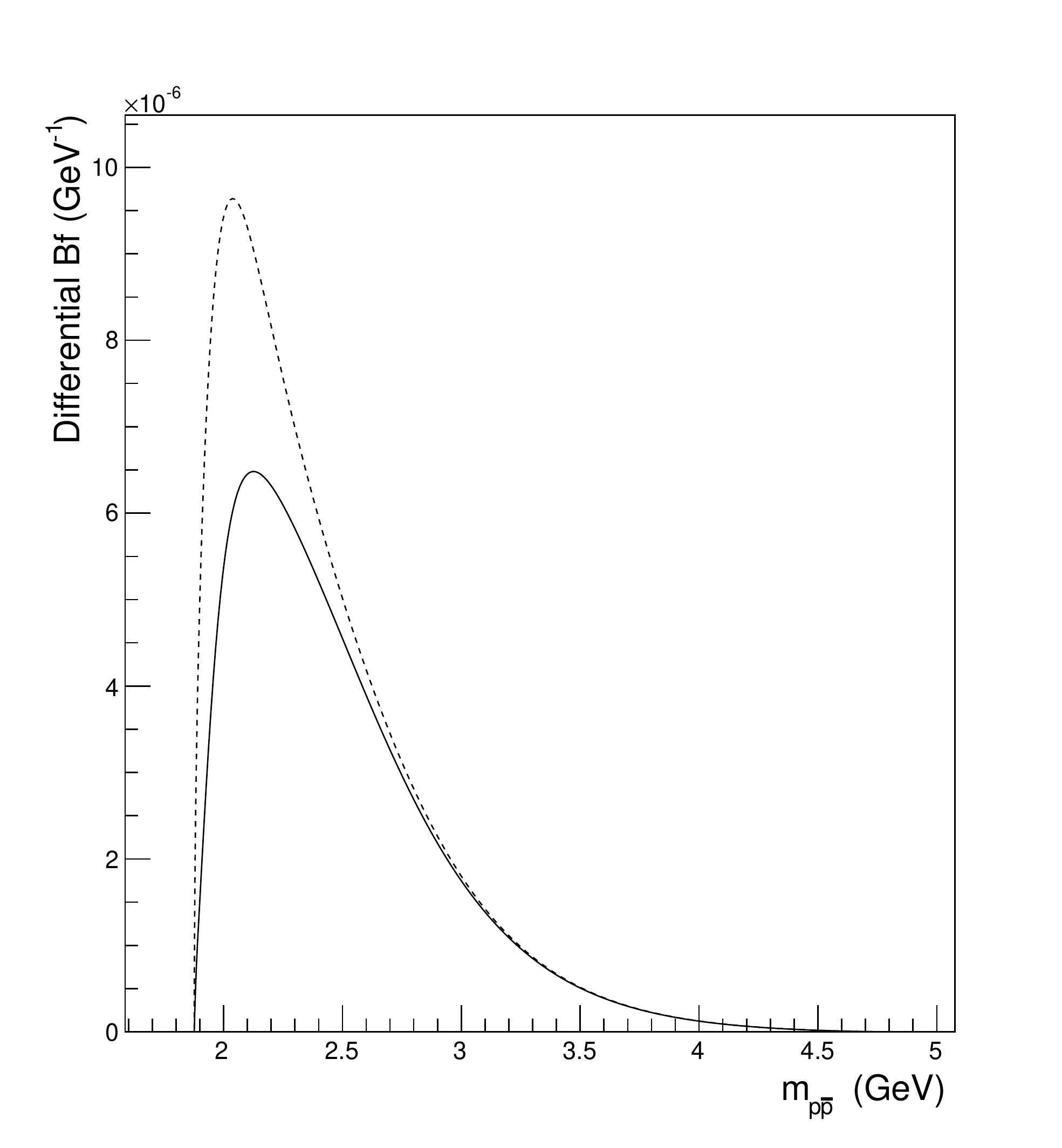}
\caption{Differential branching fractions {\it vs} $m_{p\bar p}$ for 
$B^{\pm} \to K^{\pm} p \bar{p}$ decays separately: the continuous (dashed) line refers to the $B^{\pm}$ decay.}
\label{fig:br}
\end{center}
\end{figure}

\begin{figure}[h]
\begin{center}
\includegraphics[width=0.70\textwidth] {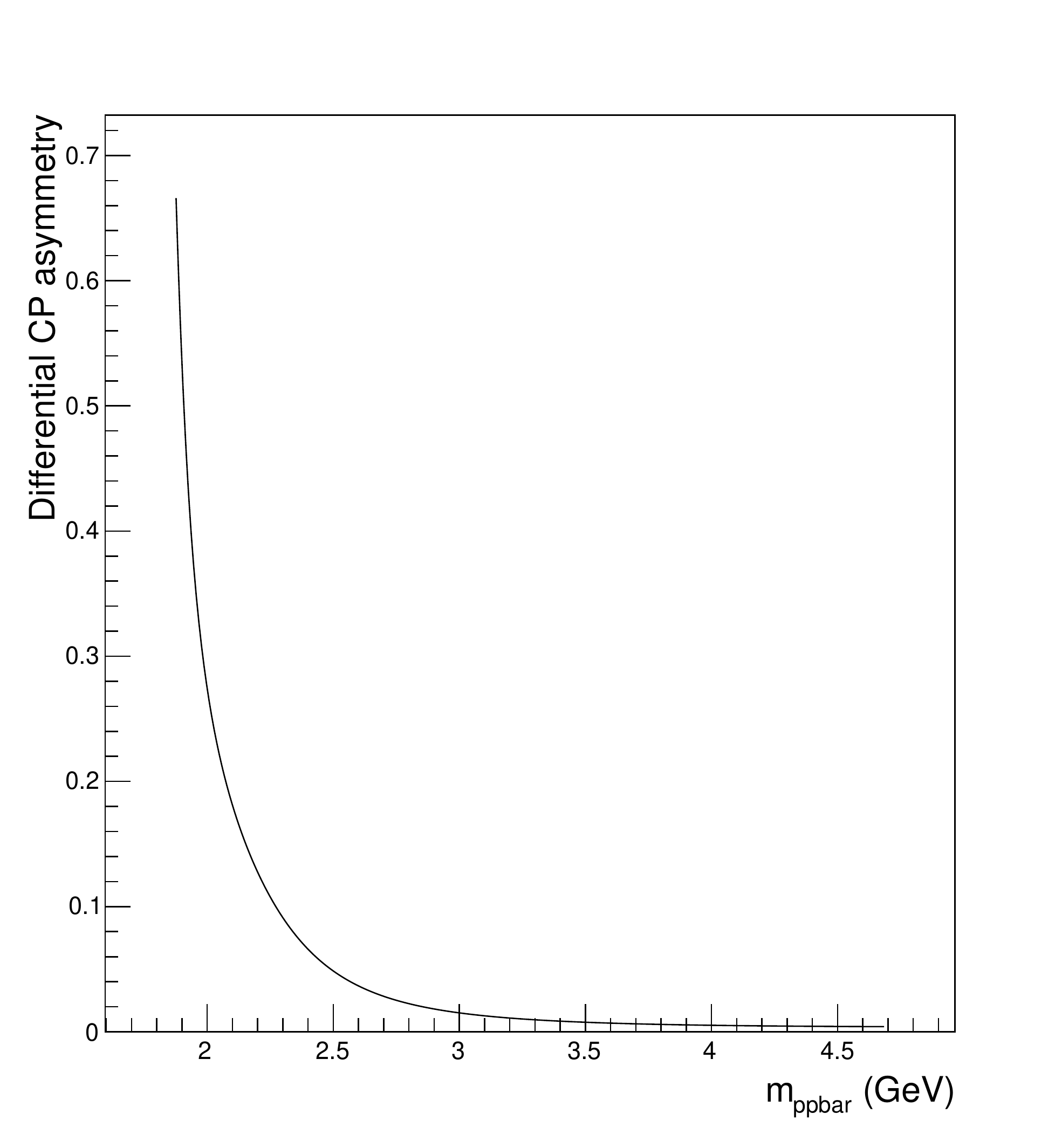}
\caption{Differential CP asymmetry {\it vs} $m_{p\bar p}$}
\label{fig:acp}
\end{center}
\end{figure}

 Table \ref{tab:two} shows the experimental values of the branching fraction and of the CP and FB asymmetries and the theoretical 
results about such quantities, both according to the factorization assumption and to our best fit. Furthermore, figs. \ref{fig:br} and \ref{fig:acp} represent, respectively, the differential branching fractions of the $K^+$ and $K^-$ decays separately and the differential 
CP-asymmetry, ${\cal A}_{CP}$, eq. (\ref{dcp}), which could be measured in the future. 

Two comments are in order about the results given in Table \ref{tab:one}
and in Table \ref{tab:two}. First of all, the parameters $C_A$ and $S_0$, as determined by our best fit (see line F2 of Table \ref{tab:one}), are affected by large, asymmetric errors; however, while the value of $C_A$ is compatible with factorization (compare with line F1), $S_0$ is much greater, which suggests that the current term derives a large contribution from the effects described in subsect. 3.2. As regards Table \ref{tab:two}, the theoretical value of $A_{CP}$ is affected by a large error, while the one of $A_{FB}$ is quite small. This is connected to the experimental errors and to the definition itself of these asymmetries. $A_{CP}$ is proportional to the difference between two large and almost equal quantities and is known experimentally with a relative error of $141\%$; therefore, its dependence on the fit parameters is quite mild.
The opposite happens for $A_{FB}$, whose experimental value is not so close to the theoretical one and is affected by a quite small relative error\cite{lhcb}.

\begin{table}[h!]
\centering
\setlength{\tabcolsep}{1.5pc}
\catcode`?=\active \def?{\kern\digitwidth}
\caption{Branching fraction and CP and FB asymmetries: experimental data (Ex) 
are compared to results of the factorization assumption (F1) 
and of our fit (F2). $Bf$ is multiplied by $10^6$, $A_{CP}$ and $A_{FB}$ by $10^2$. Here the experimental errors of the asymmetries 
are summed quadratically (see eqs. (\ref{bf-}) and (\ref{dbf})).
}

\label{tab:two}
 
\begin{tabular}{rlll}
\hline
Fit &     $~~~~~Bf$   &     $~~~A_{CP}$     &  $~A_{FB}$   \\
\hline 
Ex    & $ 5.9\  \pm 0.5        $   &  $ -2.2 \pm 3.2~~~$   &   $ 37.0\pm 2.4  $\\[5pt]
F1    & $ 2.96 \pm 0.27  $  & $\ \  8.0^{\ +0.24}_{\ -0.29}  $   &  $ 7.6\pm 0.1  $ \\[5pt]      
F2    & $ 6.21^{\ +2.40}_{\ -0.49}  $  & $ ~11.7^{\ +9.5}_{\ -7.1}  $   &  $ 11.4_{-1.5}^{+0.1} $ \\[5pt]
\hline
\end{tabular}
\end{table}
%
%



\section{Discussion and Conclusions}

Here we comment on our main results.

a)  As regards the current contribution, we have adopted the parametrization suggested by the pole model\cite{mvs}, like ref. \cite{cht}. As already observed in subsect. 3.4, by comparing the lines L1 and L2 of Table \ref{tab:one}, we conclude that, according to our fit, the current term is much greater than expected from factorization. Moreover, we see from line F2 that $|S_0| >> |S_1|$, therefore the term corresponding to the angular momentum $J$ = 0 prevails neatly over the one with $J$ = 1, in agreement with the low-mass character of the $p\bar{p}$ peak. However, the term with $J$ = 1 gives an essential contribution to the fit to the Dalitz plot distribution, as expected from C and P violation in weak decays\cite{hms}. 

b) The FB asymmetry ($A_{FB}$) - confirmed by the Dalitz plot distribution\cite{lhcb,bb2} - is caused by the linear terms in cos $\theta_p$, which appear in the differential decay width (\ref{ddw}), precisely,

- in the Jacobian (\ref{jcb}), owing to the Lorentz boost from the $p\bar{p}$ frame to the $B$ rest frame;

- in the interference term between the pseudoscalar and the vector amplitude, whose coefficient is $\Omega_{13}$,
eq. (\ref{ifl}).   

In our model - for which, incidentally, 
the second cause of asymmetry shown above is ruled out, owing to the absence of the pseudoscalar amplitude - $A_{FB}$ is positive for the $K$ decay, not only as a result of our fit, but also according to the factorization assumption, as shown in Table \ref{tab:two}. In this connection, it is worth observing that, in the model proposed, the sign of $A_{FB}$ is not {\it a priori} determined; it depends on the moduli and phases of the amplitudes involved in the decay, according to the $\beta$ coefficients (\ref{bt1}) to (\ref{bt4}). Furthermore, this is not typical of the $K$ decay. Indeed, also in the case of $\pi$ decay, the parametrization by ref. \cite{cht} yields a positive $A_{FB}$, whereas experiments find a negative value\cite{bb2,be2,lhcb}. Therefore, our picture seems to contradict the pole model conclusions and naive quark model expectations\cite{chg}. Rather, as observed in subsect. 3.1, the negative FB asymmetry of the $\pi$ decay could be explained by means of the 
color-suppressed tree diagram.

c) In the $K$ decay, the gap between the experimental value of $A_{FB}$ and the result 
of our fit (see Table \ref{tab:two}) could be filled by the first term of eq. (\ref{bt3}), concerning the electric form factor 
of the proton, neglected also in ref. \cite{cht}, which is linear in cos $\theta_p$. This explanation of the 
large asymmetry could be alternative to the one proposed by other authors\cite{gg1}.

d) Lastly, the CP asymmetry ($A_{CP}$) is found to be positive according to our parametrization, which does not differ so much from the prediction of the factorization assumption (see also ref. \cite{gg1}); on the contrary, experiments seem to indicate a negative value, although compatible with a positive asymmetry within errors. Therefore, if a more accurate determination of this observable could be desirable, also a deeper analysis of the model is needed. Indeed, although less important than in meson $B$ decays, the FSI could produce a relative phase between the current term and the transition amplitude, so as to change the sign of $A_{CP}$\cite{ccs,ccs1}. In other words, the difference of sign between data and model predictions does not necessarily imply new physics.

\vskip 0.25in
\centerline{\bf Acknowledgments}
The authors are grateful to Prof. Chua for his very kind help. Moreover they thank their friends C. Patrignani and P. Saracco for useful and stimulating discussions and for technical support.

\vskip 0.40in

 \setcounter{equation}{0}
 \renewcommand\theequation{A. \arabic{equation}}

 \appendix{\large \bf Appendix}

 Here we give the numerical values of the CKM matrix elements and of the Wilson coefficients involved in the decays considered, as well as the expressions of the matrix elements that appear in the expressions of the current and of the transition terms, according to refs. \cite{cht,pdg}.
 
 \vskip 0.15in
 \subsection*{\bf CKM Matrix Elements}
 \vskip 0.15in

 We give the CKM matrix elements\cite{pdg} which enter the expressions of the decay amplitudes
 for $B^- \to K^-(\pi^-) p \bar{p}$: 

 \begin{eqnarray}
 V_{ub} &=& A \lambda^3 (\rho-i\eta), \ ~~~~ \  V_{tb} = 1, \ ~~~~ \ 
 V^*_{us} =  \lambda,
 \\ 
 V^*_{ts} &=& -A \lambda^2, \ ~~~~ \ V^*_{ud} = 1-\frac{1}{2}\lambda^2, \ ~~~~ \ V_{td}^* = A \lambda^3 (1-\rho+i\eta), 
 \end{eqnarray}
 with $A$ = 0.814, $\lambda$ = 0.226, $\rho$ = 0.117 and $\eta$ = 0.353. For the CP-conjugated decays one has to take the complex conjugated elements.

 \vskip 0.15in
 \subsection*{\bf Wilson Coefficients}
 \vskip 0.15in
 For the $B^{\pm} \to K^{\pm} p \bar{p}$ decay, we have\cite{cht}
 \begin{eqnarray}
 a_1 &=& 1.05, \ ~~~~ \ a_2 = 0.02, \ ~~~~ \ \ ~~~ \ a_3 = (72.7-0.3i) \cdot 10^{-4}, \nonumber
 \\
 a_4 &=& -(3.873+1.21i) \cdot 10^{-2},  \ ~~~~ \ a_5 = -(66+0.3i) \cdot 10^{-4}, \nonumber
 \\
 a_6 &=& -(5.553+1.21i) \cdot 10^{-2}, \ ~~~~ \ a_9 = -(92.6+2.7i) \cdot 10^{-4}. 
 \end{eqnarray}
 As for the $B^{\pm} \to \pi^{\pm} p \bar{p}$ decay, the coefficients read as
 \begin{eqnarray}
 a_1 &=& 1.05, \ ~~~~ \ a_2 = 0.02, \ ~~~~ \ \ ~~~~ \ a_3 = (73+0.3i) \cdot 
10^{-4}, \nonumber
 \\
 a_4 &=& -(3.757+1.083i) \cdot 10^{-2}, \ ~~~~ \ a_5 = (-66+0.3i) \cdot 10^{-4}, \nonumber
 \\
 a_6 &=& -(5.447+1.083i) \cdot 10^{-2}, \ ~~~~ \ a_9 = -(92.4+2.5i) \cdot 10^{-4}. 
 \end{eqnarray}

 \vskip 0.15in
 \subsection*{\bf Non-perturbative Matrix Elements}
 \vskip 0.15in

 A) The matrix element $\langle K^-|L_{sb}^{\mu}|B^-\rangle$, which appears in the current 
 term (\ref{I1}), reads as\cite{cht}
 \begin{equation}
 \langle K^-|L_{sb}^{\mu}|B^-\rangle = k^{\mu} r F_0 + l^{\mu} F_1,
 \label{mel1}
 \end{equation}
 with
 \begin{equation}
 k = p_p + p_{\bar p}, \ ~~~~ \ l = 2p_B -(1+r)k, \ ~~~~ \ r = 
 \frac{m_B^2-m_K^2}{k^2} 
 \end{equation}
 and 
 \begin{equation}
 F_0 = \frac{S_0}{1 - u_1\tau + u_2\tau^2}, \ ~~~~ \ 
 F_1 = \frac{S_1}{(1-\tau)(1 - w\tau)}, \ ~~~~ \ 
 \tau = \frac{t}{m_V^2}. \label{ffc}
 \end{equation}
 In order to get the expression of $\langle \pi^-|L_{sb}^{\mu}|B^-\rangle$,
 we have to replace $m_K$ by $m_{\pi}$ in $r$. Moreover, the numerical 
 values of the parameters which appear in eqs. (\ref{ffc}) are
 \begin{equation}
 m_V = 5.42 ~ (5.32)~ GeV, \ ~~~~ \ u_1 = 0.70~ (0.76), ~~
 u_2 = 0.27 ~ (0.28), ~~ w = 0.43 ~ (0.48),
 \end{equation}
 the values outside (inside) parentheses referring to the $K(\pi)$ decay. Lastly, $S_0$ and $S_1$, which have the dimension of an energy, are treated as free parameters and determined according to the two possible choices described in the text: see Table \ref{tab:one}.
 
 B) Now we give the expressions of other matrix elements involved in the current terms in $K$ decay\cite{cht}:
 \begin{eqnarray}
 \langle K|\tilde{L}_{sb}|B\rangle &=& \frac{m_B^2-m_K^2}{m_b-m_s} F_0,
 \\
 \langle p \bar{p}|L_{\mu}^i|0\rangle &=& \bar{u}(p_p)[\Phi^i\gamma_{\mu}+
 \hat{\Phi}^i\frac{\delta_{\mu}}{2m_p}-(g_A^i\gamma_{\mu}\gamma_5+
 \gamma_5\frac{h_A^i}{2mp}k_{\mu})]v(p_{\bar{p}}),
 \\
 \langle p \bar{p}|R_{\mu}|0\rangle &=& \bar{u}(p_p)[\Phi^2\gamma_{\mu}+
 \hat{\Phi}^2\frac{\delta_{\mu}}{2m_p}-(g_A^2\gamma_{\mu}\gamma_5+
 \gamma_5\frac{h_A^2}{2mp}k_{\mu})]v(p_{\bar{p}}),
 \\
 \langle p\bar{p}|\tilde{R}_{ss}|0\rangle &=&
 (\frac{m_p}{m_s}g_A^3+\frac{k^2}
 {4m_p m_s}h_A^3) \bar{u}(p_p) \gamma_5 v(p_{\bar{p}}).
 \end{eqnarray}
 
 Here the first and last formulae are deduced from equations of motion; in
 particular, the first one exploits also eq. (\ref{mel1}), contracted with 
 $k_{\mu}$, and the relation $k\cdot l$ = 0. Moreover

 \begin{equation}
 \delta = p_{\bar p} - p_p,
 \end{equation}

 \begin{eqnarray}
 \Phi^1 &=& G_M^p - G_M^n + S_V, \ ~~~~ \ \Phi^2 = 3(S_V-G_M^n), \ ~~~~ \
\ ~~~~ \ \nonumber
 \\ 
 \Phi^3 &=& S_V-G_M^p - 2G_M^n, \ ~~~~ \  \Phi^4 = G_M^p, \ ~~~~ \ 
\ ~~~~ \ \ ~~~~ \ \label{ffe}
 \\
 G_M^p &=& \lambda(t)\sum_{n=1}^5\frac{x_n}{t^{n+1}}, \ ~~~~ \ G_M^n = 
-\lambda(t)\sum_{n=1}^2\frac{y_n}{t^{n+1}}, \ ~~~~ \ S_V = -\lambda(t)\frac{y_3}{t^2},   
 \\
 \lambda(t) &=& [ln(t/\Lambda_0^2)]^{\gamma}, \ ~~~~ \ \Lambda_0 =0.3 ~ GeV, \     
 ~~~~ \ \gamma = -2.148,
 \\
 x_1 &=& 420.96 ~ GeV^4, \ ~~~~ \  x_2 = -10485.50 ~ GeV^6,\ ~~~~ \  
 \ ~~~~ \ \nonumber
 \\
 x_3 &=& 106390.97 ~ GeV^8, \ ~~~~ \  x_4 = -433916.61 ~ GeV^{10},
 \\ 
 x_5 &=& 613780.15 ~ GeV^{12}, \ ~~~~ \ \ ~~~~ \ \ ~~~~ \
 \\
 y_1 &=& 236.69 ~ GeV^4  \ ~~~~ \  y_2 = -579.51 ~ GeV^6, \ ~~~~ \  y_3 = -52.42 ~ GeV^4  
 \end{eqnarray}
 and
 \begin{eqnarray}
 g_A^1 &=& f_A + d_A + s_A, \ ~~~~ \ \ g_A^2 = 2d_A + 3s_A, \ ~~~~ \ 
\ ~~~~ \ \nonumber
\\
 g_A^3 &=& -f_A + d_A + s_A, \ ~~~~ \  \ g_A^4 = f_A + 1/3 d_A, 
\ ~~~~ \ \ ~~~~ \
 \\
 f_A &=& \lambda(t)\sum_{n=1}^2\frac{w_n}{t^{n+1}}, \ ~~~~ \ d_A = \lambda(t)
 \sum_{n=1}^2\frac{z_n}{t^{n+1}}, \ ~~~~ \ s_A = \lambda(t)\frac{z_3}{t^2},   
 \\
 w_1 &=& 399 ~ GeV^4, \ ~~~~ \ w_2 = -1055 ~ GeV^6, \ ~~~~ \ \ ~~~~ \ \ ~~~~ \
 \\
 z_1 &=& 65.93 ~ GeV^4, \ ~~~~ \  z_2 = -1055 ~ GeV^6, \ ~~~~ \  z_3 = 333.06 ~ GeV^4.
 \end{eqnarray}
 Lastly, we have set 
\begin{equation}
\hat{\Phi}^i = 0 ~~~~ {\mathrm and} ~~~~ h_A^i = \frac{-4m_p^2}{t-m_{\pi_0}} g_A^i, 
 ~~~~ i = 1 ~~ {\mathrm to} ~~ 4. \label{fff}
\end{equation} 
 The corresponding expressions for the $\pi$ decay are obtained
 by replacing $m_K$ by $m_{\pi}$, $m_s$ by $m_d$, $\Phi^3$ by $S_V$ and  
 $g_A^3$ by $s_A$.

 C) The matrix elements involved in the transition term (\ref{trt}) are\cite{cht}
 \begin{eqnarray}
 \langle K|\tilde{L}_{us}|0\rangle \langle p \bar{p}|S^{ub}|0\rangle &=& 
 -f_K m_b [F_A \bar{u}(p_p) p\hspace{-0.45 em}/_K\gamma_5 v(p_{\bar{p}} +
  F_P\bar{u}(p_p) \gamma_5 v(p_{\bar{p}})],
 \\
 \langle K|\tilde{L}_{us}|0\rangle \langle p \bar{p}|S^{ub}|0\rangle &=& 
 -f_K m_b [F_{V5} \bar{u}(p_p) p\hspace{-0.45 em}/_K v(p_{\bar{p}} +
  F_P\bar{u}(p_p) v(p_{\bar{p}})]. 
 \end{eqnarray}
 Here $f_K$ = 0.158 $GeV$ and 
 \begin{equation}
 F_A = \frac{C_A}{t^3}, \ ~~~~ \ F_{V5} = \frac{C_{V5}}{t^3}, \ ~~~~ \ 
 F_P = \frac{C_P}{t^4}, \label{ffg}
 \end{equation}
 $C_A$, $C_{V5}$ and $C_P$ being assumed, again, as free parameters and determined as described in the text.

 In the case of the $\pi$ decay, one has to replace $f_K$ by
 $f_{\pi}$ = 0.133 GeV.


\end{document}